\documentclass[aps,prl,showpacs,twocolumn,groupedaddress]{revtex4-1}
\usepackage{graphicx,graphics,color,epsfig}
\usepackage{amsmath}
\usepackage{amssymb}
\usepackage{amsfonts}
\usepackage{amsfonts}
\usepackage{epstopdf}
\usepackage{bm}
\usepackage{times,xspace}
\usepackage{color}

\begin{document}


\title{Quantum oscillations from `open' Fermi surface in quasi-one-dimensional lattices: Application to YBa$_2$Cu$_3$O$_{6+x}$ cuprates, organic salts, ladder compounds, and related systems}

\author{Pranoy S. Varma$^1$, and Tanmoy Das$^2$}
\affiliation{$^1$Department of Electrical Engineering, Indian Institute of Technology Madras, Chennai-600036, India,\\ $^2$Department of Physics, Indian Institute of Science, Bangalore-560012, India}

\date{\today}
\begin{abstract}
According to the celebrated Onsagar-Lifshitz paradigm, the observation of Shubnikov de-Haas and de-Haas van Alphen (SdHvA) oscillations is an indication of the presence of `closed' orbit Fermi surface in the bulk. We present a real-space based calculation of SdHvA oscillations in generalized quasi-one-dimensional lattices by relaxing the quasi-classical approximations embedded in this decades old Onsagar-Lifshitz paradigm. We find that sizable quantum oscillation can arise from `open' Fermi surfaces as long as
cyclotron orbits can form in real-space with finite, but not necessarily equal, electron hopping along both $x$- and $y$-directions. Our results quantitatively explain the puzzling emergence of SdHvA oscillation in various quasi-one-dimensional materials, including the chain state of YBa$_2$Cu$_3$O$_{6+x}$ cuprates, organic materials, various ladder compounds, weakly coupled linear chains, or quantum wires, and other related systems.
\end{abstract}
\pacs{71.18.+y,72.15.Gd, 74.25.F-, 74.72.Gh}
\maketitle

Shubnikov de-Haas and de-Haas van Alphen (SdHvA) oscillations, as often called quantum oscillation (QO), are widely studied measurements to probe the bulk Fermi surface (FS). According to the widely used Onsagar-Lifshitz paradigm,\cite{Onsagar,Lifshitz,Shoenberg} which is based on the quasi-classical quantization of the real-momentum phase space, QO is directly proportional to the FS area: $F=\Phi_0/2\pi^2S_{k}$, where $\Phi_0=h/e$ is the flux quanta, and $S_k$ is the cross-sectional area of the FS normal to the applied field direction. Due to this relationship, the observation of QO is often attributed as a proof to the presence of the `closed' FS orbit in the bulk electronic structure. However, a number of recent observations has challenged this scenario. For example, QO is recently observed in a Kondo insulator.\cite{SmB6} QO is also observed in several quasi-one-dimensional (1D) organic salts (with even solely open FS),\cite{OrgNamReview,OrgHouse,OrgSasaki,OrgCaulfield,OrgDannerFS,OrgDannerExp} and in weakly coupled linear chains\cite{QOlinearchain}. The Hall-effect data of quasi-1D ladder cuprates Sr$_{14-x}$Ca$_x$Cu$_{24}$O$_{41}$ [Ref.~\onlinecite{ladderHall}] PrBa$_2$Cu$_4$O$_8$ [Ref.~\onlinecite{ladderHorii}], in other ladder compounds\cite{LiMoOHall} are equally puzzling due to the presence of open orbit FS in these systems. 

\begin{figure}[t]
\hspace{2cm}
\includegraphics[width=3.5in]{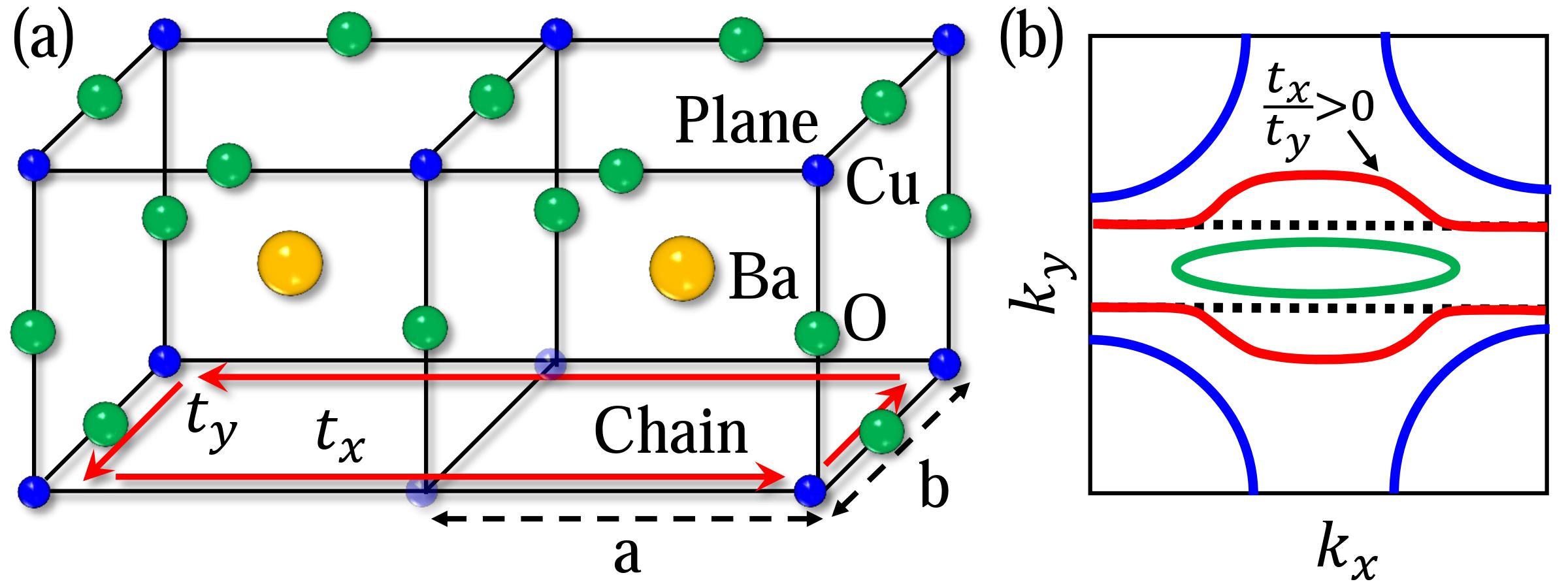}
\caption{(Color online) (a) A typical schematic Y123 lattice with an isotropic CuO$_2$ plane (top layer) and a quasi-1D CuO chain (bottom) layer in the Ortho-II phase. Due to missing in-plane O atoms in the chain layer, every either Cu atoms become inert, making an effective rectangular chain lattice.\cite{StrAndersen,StrZimmermann} Horizontal red arrows dictate a possible close path for the electron trajectory with anisotropic hopping. (b) Schematic FS for four representative cases: pure 1D FS for $t_x=0$ (black dashed), warped open FS for $t_x/t_y>0$ (red solid), and two closed FSs  with electron- (green) and hole- (blue) pockets.}
\label{fig1}
\end{figure}

YBa$_2$Cu$_3$O$_{6+x}$ (Y123),\cite{QOLeyraud,QOSingleton,sebastian_compensated,sebastianmass,QOreview} YBa$_2$Cu$_4$O$_8$ (Y124)\cite{QOYelland} cuprates add to this puzzle with peculiar results. YBCO crystal exhibits drastic structural transition from the disorder tetragonal phase in the underdoped region to the so-called superstructure Ortho-II phase in the doping range of $0.3<x<0.67$, i.e. 6.3 to 6.67 oxygen content in which a CuO-chain layer forms with missing oxygen atom in alternating chains (see Fig.~\ref{fig1}(a)).\cite{StrAndersen,StrZimmermann}  
In the Ortho-II phase, series of high magnetic field measurements have demonstrated peculiar transport properties, which are not consistent with the spectroscopic features measured at zero magnetic field. Hall-effect, supported by other transport\cite{seebeck,seebeckNC} measurements have observed small and negative Hall resistance which is indicative of electron-like FS.\cite{YBCOelectron_pocket} Moreover, numerous QO data\cite{QOLeyraud,QOSingleton,sebastian_compensated,sebastianmass,QOreview,QOYelland} exhibited the emergence of oscillation with small frequency (530~T) in this doping range. According to the Onsagar-Lifshitz paradigm, these measurements suggest the presence of small electron-like FS pocket, occupying only 2\% area of the FS, which is significantly lower than the nominal doping concentration. Moreover, the QO frequency remains doping independent throughout the doping region it is observed, and vanishes sharply away from the doping region where Ortho-II structural phase also disappears.\cite{sebastianmass,QOreview} Such electron pocket does neither arise naturally from the band-structure calculations considering the CuO$_2$ planes, nor seen in the spectroscopic data\cite{ARPES_Damascelli,ARPES_Borisenko,STM_FA} Various density wave formalisms predict the formation of doping dependent electron-pockets,\cite{DDW,Norman,Kivelson,SachdevSDW,DasEP,SebastianCDW} which are yet not detected by angle-resolved photoemission spectroscopy (ARPES),\cite{ARPES_Damascelli,ARPES_Borisenko} and scanning tunneling spectroscopy (STS) data.\cite{STM_FA}

\begin{figure}[t]
\hspace{2cm}
\includegraphics[width=3.5in]{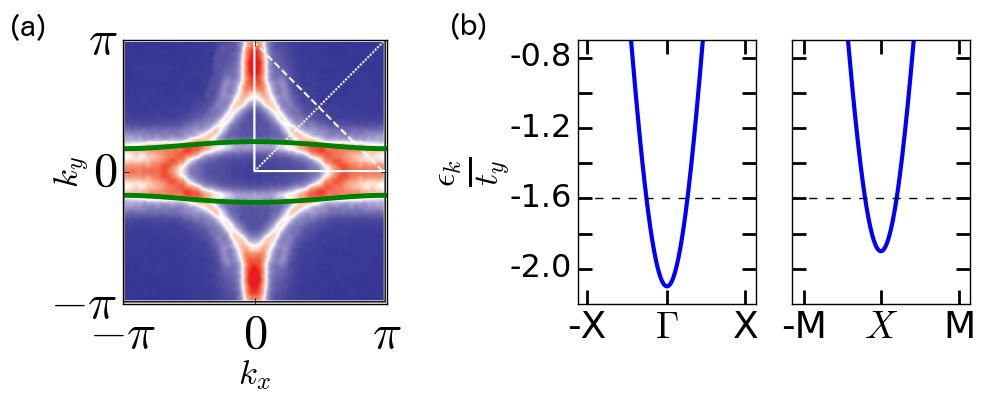}
\caption{(Color online) (a) Fitting of the calculated FS (green lines) with the chain state of Y123 from ARPES data\cite{ARPES_Damascelli} at $x=0.29$. (b) The corresponding band dispersion is shown along two representative directions to demonstrate that  the chain states are electron like.}
\label{fig2}
\end{figure}

Based on these anomalous appearances of QO in variety of systems, we explore a QO calculation for systems with open FS. But in the corresponding real-space, the electron trajectory encloses a close path with anisotropic hoppings, as shown in Fig.~\ref{fig1}(a). In particular we take the case of an open FS, but warped along the perpendicular direction for a quasi-1D system, as demonstrated in Fig.~\ref{fig1}(b). Such FS corresponds to a rectangular lattice, or two weakly connected atomic chains or quantum-wires in which the electron hopping along the direction of the chain (say $t_y$) is different than that between the chains (say $t_x$). As long as $t_x>0$, the electron is allowed to form a closed cyclotron orbit with the application of magnetic field, without necessarily commencing a closed FS in the momentum space. For $t_x=0$, the FS consists of two disconnected 1D line, and the QO is forbidden even in the real-space picture. As $t_x$ increases, the FSs become warped along the $k_y$ directions [red line in Fig.~\ref{fig1}(b)]. In this case, even though the two FS lines are not adiabatically linked, the quantum tunneling of electrons between the two chains is active with limited $k_x$-values. Interestingly, as long as $t_x$ and $t_y$ have the same sign, the corresponding FS topology for $t_x/t_y<<1$ is electron-like, centering the $\Gamma$-point (see Fig.~\ref{fig2}). As $t_x/t_y$ increases, values of the Fermi momenta either reduce or increase depending on the chemical potential. Above a critical value, FS becomes closed, forming either an electron- [green line in Fig.~\ref{fig1}(b)] or hole-pocket [blue line in Fig.~\ref{fig1}(b)].

Our theoretical calculations have two parts. In the first part, we solve a lattice model in which magnetic field is included within the Peierls substitution, and the quantization condition is imposed via the quantization of the magnetic flux. Our main result is that QO can arise as long as $t_x>0$ with open FS. In the second part, we substantiated the results with a low-energy continuum model with anisotropic band mass (effectively modeling the hopping anisotropy of the rectangular lattice), and the magnetic field is employed with a standard Landau gauge. In this case, we find an interesting result that the charge density is a more generic quantity that dictates the QO frequency, and it becomes equal to the area of the closed FS pocket (within Luttinger theorem) as predicted by the Onsagar-Lifshitz theorem.\cite{Onsagar,Lifshitz,Shoenberg} We present the evolution of the QO in the magnetization profile as a function of magnetic field as $t_x/t_y$ is varied while keeping the carrier density constant, and vice versa. We find an interesting result that the oscillation frequency depends weakly on the FS warping factor (i.e., $t_x/t_y$) and sharply reduces as the FS undergoes transition from the open to closed orbit topology. Consistently, since in Y123 sample, the O doping effect does not change the warping effect (i.e, the $t_x/t_y$ ratio), the corresponding QO frequency remains unchanged with doping.\cite{sebastianmass,QOreview}   

{\it Model:} We study a realistic single band tight-binding model in a rectangular lattice with nearest neighbor hopping elements. In the absence of magnetic field, the non-interacting dispersion in the momentum space takes the form of $\xi_{\bf k}=-2t_x\cos{(k_xa)}-2t_y\cos{(k_yb)}-\mu$, where $\mu$ is the chemical potential. We obtain the corresponding parameters by fitting the chain state measured by ARPES\cite{ARPES_Damascelli,ARPES_Borisenko} for Y123 at $x=$0.29, as $t_x/t_y=$0.05 and $\mu=-1.6t_y$, see Fig.~\ref{fig2}(a). The parameter values remain the same for fitting the ARPES data at other dopings, since the chain does not change with doping.\cite{ARPES_Damascelli,LDA,LDA2}  The corresponding band dispersion plotted in Fig.~\ref{fig2}(b) shows that the band bottom lies below the $\Gamma$ point, suggesting that the quasiparticles on the chain FS consist of electrons. 

We consider the case where the magnetic field ($B$) is oriented perpendicular to the CuO chain. For convenience, we take a Landau gauge as ${\bf A}=Bx{\hat y}$. This particular choice breaks the translation symmetry along the $x$-axis, while respects it along the $y$-axis. In this case, a new translational symmetry can be imposed with an appropriate gauge transformation by defining a magnetic translational operator $\tau_{{R}}  \rightarrow T_{{R}}\text{e}^{i \sum_j{\frac{e}{\hbar}BR_xy_j}}$, where $T_R$ is the translation operator without magnetic field, $R_x$ (= $a$) is the primitive translational vector along the $x$-direction and $j$ labels an individual electron. As shown in the supplementary material (SM),\cite{SM} the translational symmetry reemerges in both directions as the flux through a commensurate magnetic unit cell becomes a rational number of the quantum of flux ($\Phi_0$), providing the essential quantization condition $\Phi={\bf B}.({\bf a}\times{\bf b})=\frac{p}{q}\Phi_0$, where $p$ and $q$ are integers. Therefore, without loosing generality, we define the magnetic unit cell with primitive vector ${\bf R}=m(qa){\hat x}+nb{\hat y}$, where $m$ and $n$ are integers, which would allow a magnetic flux of $p\Phi_0$. 

The corresponding gauge field modifies the electron hopping by adding a phase factor (called Peierls phase) as $ t_{mn}^{m'n'}\rightarrow t_{mn}^{m'n'}\mathrm{e}^{-i\phi_{mn}^{m'n'}}$, where $\phi_{mn}^{m'n'}=\frac{e}{\hbar}\int_{mn}^{m'n'}{\bf{A}.d\bf{l}}$. For our choice of the gauge potential, the Peierls phase vanishes along the $x$-direction. For the nearest-neighbor hopping along the $y$-direction, and the Peierl phase at the $m^{\rm th}$-site becomes $\phi_{mn}^{m,n\pm1}=\pm\frac{e}{h}B(ma)b=\pm\frac{p}{q}2\pi$. Finally, employing periodicity in the magnetic Brillouin zone, we obtain the lattice model
\begin{eqnarray}
H&=&-\sum_{rs}\left[\sum_{m=0}^{q-2}t_{x}{c^\dagger}_{rs,m}{c}_{rs,m+1}-t_{x}{c^\dagger}_{rs,q-1}{c}_{rs+q,0}\right. \nonumber\\
&&~~~~~~~-\sum_{m=0}^{q-1}t_{y}\text{e}^{im\phi}{c^\dagger}_{rs,m}{c}_{rs+1,m}+{\rm h.c.}\Bigg],
\label{Hreal}
\end{eqnarray}
where $r,~s$ label the position of the magnetic unit cell, and $m$ labels the positions of atoms inside the magnetic unit cell. For a given magnetic field, the magnetic unit cell adjusts itself in such a way that the flux through it is an integer multiple of the quantum of flux, $\Phi=M_xM_yB=\frac{p}{q}\Phi_0$ as mentioned before, where $M_x$ and $M_y$ are the lengths of the magnetic unit cell respectively. By convention, we choose $M_y=2$ and $M_x=q$. Considering the magnetic Brillouin zone with $q$-number of sub-lattices, we solve the above Hamiltonian by Fourier transforming to the corresponding momentum space (see SM\cite{SM}). The magnetic field is varied by changing the length of the magnetic unit cell $q$ with a single quantum of flux per magnetic unit cell ($p=1$). The magnetization at zero temperature can be calculated easily from the total energy as $M=-\frac{\partial \epsilon_{\rm total}}{\partial B}$, where the total system energy is $\epsilon_{\rm total}=\sum_{\nu,\xi^{\nu}_{\bf k}\le\mu}\xi_{\bf k}^{\nu}$. $\xi_{\bf k}^{\nu}$ is the $\nu^{\rm th}$-eigenvalue of Hamiltonian in Eq.~\eqref{Hreal}. The SdHvA oscillations are calculated for magnetic fields with the corresponding number of sub-lattices in the magnetic unit cell less than 80 ($q<80$).

\begin{figure}[t]
\hspace{2cm}
\centering
\rotatebox{0}{\scalebox{.345}{\includegraphics{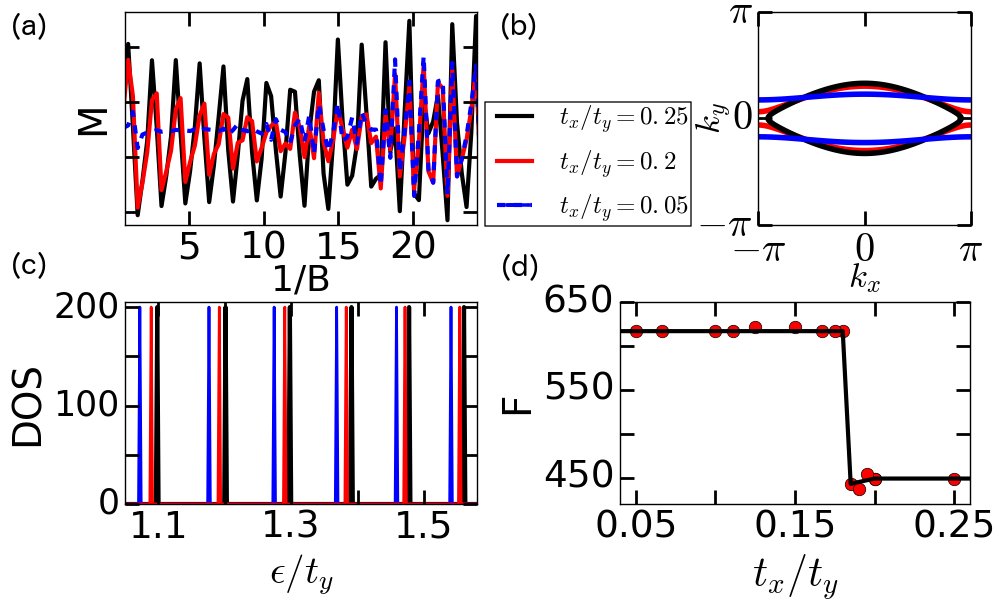}}}
\caption{(Color online) (a) Magnetization oscillation profile for the experimental parameter (blue dashed line), compared with two other values near the transition from open to closed FS topology. Both magnetic field and magnetization are in units of $t_y=1$. (b) Corresponding FS topologies for the three cases discussed in (a), plotted with the same color. (c) Oscillation in DOS at the same three parameter sets. (d) QO frequency extracted by Fourier transformation of the QO profile shown in (a). The black line is guide to the eye. For all calculations in this figure, the value of $t_x/t_y$ is changed while the carrier density is kept constant. }
\label{fig3}
\end{figure}

The calculated magnetization as a function of $1/B$ is shown in Fig.~\ref{fig3}(a) for several representative values of $t_x/t_y$. The corresponding chemical potential ($\mu$) is calculated for each $t_x/t_y$  to keep the carrier concentration unchanged. The corresponding FS topology and density of states (DOS) at the Fermi level are shown in Figs.~\ref{fig3}(b) and \ref{fig3}(c), respectively. Fig.~\ref{fig3}(d) shows the characteristic frequency of QO as a function $t_x/t_y$ with charge density remaining constant. The important result is that for the experimental open and warped FS (at $t_x/t_y=0.05$), there is a prominent oscillation in $M$ whose amplitude depends strongly on the field strength. The corresponding QO frequency is $\sim$600~T , which is close to the experimental value for this material. Expectedly, the oscillation frequency remains very much independent of $t_x/t_y$ since the carrier density is kept constant. Above a threshold value of $t_x/t_y\gtrsim 0.2$, the FS becomes almost closed and it fully closes at $\sim$0.25. In these cases, the oscillation amplitude becomes less dependent on $B$, and survives up to higher field strength. But the frequency undergoes a drastic transition to a considerably reduced value (see Fig.~\ref{fig3}(d)). On the other hand, the corresponding oscillation in the DOS does not change accordingly across this transition. This indicates that the `discontinuous' change in the frequency is not directly related to the closing of the FS topology, rather related to a sharp transition in the quantization condition. This is also evident in the field dependence of $M$ in Fig.~\ref{fig3}(a), where we see that for the open FS case, the oscillation vanishes gradually with increasing $B$, as one would expect from the transition between the quantum to classical limits. Such transition is however absent for the closed FS case. We believe that for the closed FS, the QO arises primarily from the semiclassical real-momentum phase space quantization (Bohr-Sommerfeld quantization), and all orbits possess the same size, constrained by the FS area. On the other hand, in the case of open FS, with increasing $B$, the radius of the orbits reduces strongly, and thus the oscillation amplitude also gradually vanishes.   

\begin{figure}[t]
\hspace{2cm}
\centering
\rotatebox{0}{\scalebox{.345}{\includegraphics{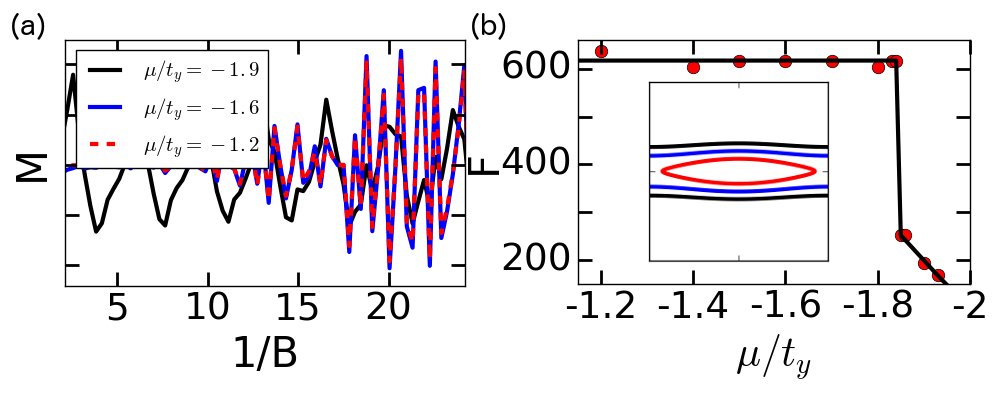}}}
\caption{(Color online) (a) Similar QO profile for $M$ for three chemical potentials (keeping $t_x/t_y$ constant) near the FS topological transitions. (b) Corresponding QO frequency dependence shows a jump in going from open to closed FS topology as in the case of $t_x/t_y$ dependence in Fig.~\ref{fig3}. {\it Inset:} FS topologies for the same parameter sets.  }
\label{fig4}
\end{figure}

Further insight to the evolution of the QO profile and frequency across the FS topological change can be obtained from the study of the carrier density dependence in Fig.~\ref{fig4}. Here we varied the chemical potential across the FS topological transition while keeping the FS warping ratio $t_x/t_y$ constant to 0.05. Here we observe a qualitatively similar trend. In the cases, the FS is open (red and blue curves), the oscillation occurs at a large value of $1/B$, and remains very much independent of $t_x/t_y$. The amplitude of the oscillation disappears gradually with higher $B$. As the FS is closed, the frequency drops by about 1/3, while amplitude becomes less sensitive to the field. These results further affirm our premise that the QO frequency for open and closed FS topologies arise from different, yet equivalent, quantization conditions.     

{\it Analytical formalism for the QO:} In order to obtain a quantitative expression for the QO frequency, we provide a Landau level calculations in the continuum model, by allowing anisotropic band masses. Under the same choice of gauge, we solve the Schr\"odinger equation $H=p_x^2/2m_x + (p_y-eBx/c)^2/2m_y$, where $m_x$ and $m_y$ are the effective band masses obtained from the same anisotropic dispersion $\xi_{\bf k}$, and other symbols have usual meanings. Since $H$ commutes only with $p_y$, we substitute $p_y=\hbar k_y$, and $p_x=-i\hbar\partial/\partial x$ to obtain a standard simple harmonic oscillator equation:
 \begin{equation}
H=-\frac{\hbar^2}{2{m}_x}\frac{\partial^2}{\partial x^2}+\frac{1}{2}{m}_y\omega_0^2\left(x-\frac{\hbar{k}_y}{eB}\right)^2
\end{equation}
The harmonic oscillator has the center of the potential at $x_0=\frac{\hbar k_y}{eB}$, and the natural frequency $\omega_0=\frac{eB}{\sqrt{m_xm_y}}$, and the corresponding eigenvalues are $E_n=\left(n+\frac{1}{2} \right)\hbar\omega_0$. Given that all allowed $x_0$ should be within the sample of dimension $L_x\times L_y$, we obtain the essential condition for magnetic field dependent degeneracy as $0<k_y<eBL_x/\hbar$. Therefore, the highest degeneracy is $D=\frac{2eBL_xL_y}{h}=\frac{2eBS}{h}$ (factor `2' arises due to spin), where $S$ is the sample's total cross-section perpendicular to the magnetic field. Let us assume that for a given magnetic field all the levels up to the $(\eta-1)^{\rm th}$ Landau level are completely filled, and the $\eta^{\rm th}$- Landau level has a fractional filling factor of $\nu$ (with $0\le\nu\le1$). Therefore, the carrier density per atom (here Cu atom) as a function of $B$ can be deduced as to be $\rho=(\eta+\nu)\frac{2eB}{h}$. By summing over all the Landau levels below the Fermi level having degeneracy $D$, we obtain the total energy as
\begin{equation}
\epsilon_{\rm total}=\frac{Sh^2\rho^2}{2\pi\sqrt{m_xm_y}}\left(1+\frac{\nu-\nu^2}{(\eta+\nu)^2}\right).
\end{equation}
Given that $\nu$ only varies between 0 to 1, the DOS and the total energy acquires oscillations as a function of $B$. This is the essential mechanism of QO for quasi-1D systems which manifests into various thermodynamical and transport properties as one enters into the quantization region. In a pure 1D-case ($m_y\rightarrow 0$), the oscillation disappears, while for a pure isotropic system ($m_x=m_y$) we can recover the earlier results.\cite{Onsagar,Lifshitz} Here we focus on studying the QO in the magnetization which turns out to be
\begin{equation}
M=\frac{2S\hbar \rho}{\sqrt{m_xm_y}}(1-2\nu),
\end{equation}
where we have neglected the higher order terms $\mathcal{O}\left(\nu^2/r^2\right)$, since usually $\eta$ is of the order of $10^8$ in real materials. Evidently, the period of the oscillation in our model depends on the carrier density as $F=2e/\rho h$. (In the case of a closed FS, $\rho$ is proportional to the FS area and thus the Onsagar relation can be recovered.)

{\it Application to cuprates:} We now examine the consistency of the theoretical results with series of data in cuprates. With O-doping in Y123, the carrier density of the CuO$_2$-plane state changes, but that of CuO does not change much as demonstrated by ARPES data.\cite{ARPES_Damascelli,LDA,LDA2} In Y123 compound, the QO frequency, observed in the Ortho-II phase in the doping range of $p\sim0.10-0.125$, varies only in the range of 530-570~T, \cite{QOLeyraud,QOSingleton,sebastian_compensated,sebastianmass}, while Y124, which has double chains, has a slightly higher frequency of 660~T.\cite{QOYelland} There are three other cuprate materials in which QO arises without the presence of chain state. In underdoped ($p\sim$0.09) single layer HgBa$_2$CuO$_{4+\delta}$ (Hg1201), QO is observed with frequency $840\pm30$~T and negative Hall sign.\cite{QOHBCO}  In this compound, however, the HgO and CuO$_2$ hybridized band produces a tiny electron pocket, centering $k=(\pi,0)$ and equivalent points as shown by band structure calculations\cite{LDAHBCO1,LDAHBCO2,HBCODas}. Therefore, it can be occupied easily with increasing magnetic field,   and can possibility give rise to QO. In the overdoped Tl$_2$Ba$_2$CuO$_{6+\delta}$ at $p=0.30$, a very large frequency QO ($F\sim$18,100~T) is observed with positive Hall sign.\cite{QOVignolle_overdoped}  This is however expected since in this doping range, the full CuO$_2$ hole pocket forms [blue line in Fig.~\ref{fig1}(b)]. In electron-doped Nd$_{2-x}$Ce$_x$CuO$_4$, a small QO $F\sim 300$~T is observed around the optimal doping region ($x=0.15$) with positive Hall sign.\cite{QONCCO} This result is however well understood due to the FS reconstruction near the antiferromagnetic critical point, stipulating tiny hole pocket at the Brillouin zone center.\cite{DasNCCO}

{\it Organics:} Our calculation is also applicable to other quasi-1D systems in which the observation of QO has remained a long standing puzzle.\cite{OrgNamReview} Organic salts, some of which are also found to be unconventional superconductors, contain quasi-1D chain states. $\alpha$-(BEDT-TTF)$_2M$Hg(SCN)$_2$ ($M$ K, NH$_4$),\cite{OrgHouse}  $\alpha$-ET$_2M$Hg(NCS)$_4$ ($M$ = K, TI, Rb)\cite{OrgSasaki,OrgCaulfield} are among the organic superconductor family members where QO has been observed even when only the quasi-1D chain state survives after a spin-density wave gap induced gapping of the other FSs. In fact,  $\alpha$-(BEDT-TTF)$_2M$Hg(SCN)$_2$ has a very similar FS topology as the Ortho-II YBCO sample, in that there is a warped chain state centering the $\Gamma$-point, and a hole-pocket centering the BZ corner.  Interestingly, the QO frequency for  $\alpha$-(BEDT-TTF)$_2$KHg(SCN)$_2$, and $\alpha$-(BEDT-TTF)$_2$NH$_4$Hg(SCN)$_2$ are around 567~T, and 670~T,\cite{OrgHouse} which are roughly the same to the values observed in underdoped YBCO samples. (TMTSF)$_2$ClO$_4$ is another interesting organic metal where only open-orbit FS is present, and QO is observed,\cite{OrgDannerFS,OrgDannerExp}  which can be fully explained without our calculation.

{\it Other quasi-1D systems:} Coupled linear chains are other examples where QO has been observed.\cite{QOlinearchain} Various quasi-1D ladder compounds, such as cuprates Sr$_{14-x}$Ca$_x$Cu$_{24}$O$_{41}$,\cite{ladderHall}  PrBa$_2$Cu$_4$O$_8$,\cite{ladderHorii} and Li$_{0.9}$Mo$_6$O$_{17}$\cite{LiMoOHall} where QO in magnetotransport can be explored.  Especially, as mentioned earlier, the Hall-effect results of these ladder compounds are particularly encouraging for the same reasons that with applied magnetic field closed electron motion can occur in real space even with open FS topology.\cite{DasChain} Finally, quasi-1D quantum wires of various nature are routinely grown nowadays with enormous materials flexibility, in which  QO with open FS can further be explored and tuned desirably.

{\it Conclusions:} The essential conclusion of the present work is that the observation of QO is not always an indication for the presence of `close orbit' FS in the bulk. Given that electron trajectory is required to commence closed path in real-space, for sufficiently anisotropic systems, QO can appear with open FS. Our theory is generic and helps bypass the approximations embedded in the decades old Onsagar-Lifshitz theory of  QO. With growing evidence of anomalous QO, especially in a number of quasi-1D systems with open FSs, our work will lead to a consistent explanation to them. Our theory to the puzzling appearance of small QO with electron-like quasiparticle due to electron-like chain state can be easily verified in number of ways. Dilute disorder is known to destroy pseudogap.\cite{Alloul} In such sample, any possible pseudogap related electron-pocket can be removed, and thus it would provide an ideal system to verify the possibility of the chain state induced QO. Within our theory we expect that the oscillation frequency would ideally be independent of temperate as well as doping. While the latter result is consistent with existing data, the former can be explored in future experiments.

\begin{acknowledgments}
The work is supported by the grant from the Indian Institute of Science, and facilitated by the Bardeen cluster. 
\end{acknowledgments}

\section{Lattice Model}
We consider a rectangular lattice with one atom per unit cell. The corresponding Hamiltonian is 
\begin{equation}
{H}=\sum_{m,n,m^\prime,n^\prime}-t_{m,n}^{m^\prime,n^\prime}c_{mn}^{\dagger}c_{m^\prime n^\prime},
\end{equation}
where $(m,n)$ and  $(m^\prime,n^\prime)$ are the site indices inside the unit cell, and $t$ is the tight-binding gopping amplitudes. 

{\it Inclusion of magnetic field.} In the presence of a magnetic field the Hamiltonian for a system of electrons in a periodic potential $U(x_i)$ is modified by the canonical replacement of the momentum operator as
\begin{equation}
{H}=\sum_{i}\left({\frac{{({\bf{p}_i}-q{\bf{A})}}^2}{2{m}^*}}+{U({\bf{x}}_i)}\right).
\end{equation}
This Hamiltonian is no longer lattice translational invariant. We work in the gauge ${\bf{{A}}}={{B}_0x}\widehat{y}$. Under a lattice translation ($T_{\bf{R}}$), there is an extra phase due to the vector potential
\begin{equation}
T_{\bf{R}}^{\dagger}{H}T_{\bf{R}}=\sum_{i}\left({\frac{{({\bf{p}_i}-q{\bf{A}}-eBR_x\hat{y})}^2}{2{m^*}}}+{U(\bf{x_i})}\right).
\end{equation}
To counteract this phase we introduce a unitary transformation,
\begin{eqnarray}
&&\left(\text{e}^{-\frac{i}{\hbar}\sum_{i}eBR_xy_i}T_{{\bf{R}}}^{\dagger}{H}T_{{\bf{R}}}\text{e}^{\frac{i}{\hbar}\sum_{i}eBR_xy_i}\right)= \nonumber\\
&& ~~~~~~~~~~~~~~~~~~~~~~~~~~~~\sum_{i}\left({\frac{{({\bf{p}_i}-q{\bf{A}})}^2}{2{m^*}}}+{U(\bf{x_i})}\right).
\end{eqnarray}
This operator combined with the translation operator is defined to be the magnetic translation operator for this choice of gauge :  
\begin{equation}
\tau_{\bf{R}}=T_{\bf{R}}\text{e}^{i \sum_j{\frac{e}{\hbar}BR_xy_j}}.
\end{equation}
The magnetic translation operators do not always commute with each other:
\begin{equation}
\tau_{\bf{a}}\tau_{\bf{b}}=\tau_{\bf{b}}\tau_{\bf{a}}e^{\frac{2\pi i}{h}e{\bf{B}}.({\bf{a}}\times{\bf{b}})}=\tau_{{\bf{b}}}\tau_{{\bf{a}}}e^{\frac{2\pi i}{h}e\Phi},
\end{equation}
where $\bf{a}$ and $\bf{b}$ are the primitive lattice vectors.  
We consider only those magnetic fields that have a rational number multiple of the quantum of flux through the unit cell, $\Phi={\bf{B}}.({\bf{a}}\times{\bf{b}})=\frac{p}{q}\Phi_0$. Imposing this condition, we now consider an enlarged unit cell, ${\bf{R}}=m(qa)\hat{x}+n(b)\hat{y}$ such that the flux through this supercell is an integer multiple of the quantum of flux $\Phi_0$. Thus in the presence of a magnetic field, we deal with a magnetic unit cell and the corresponding magnetic Brillouin zone defined by
\begin{equation}
0\leq k_x \leq \frac{2\pi}{qa} \text{ and } 0\leq k_y \leq \frac{2\pi}{b},
\end{equation}
with the number of discrete points depending on the choice of sample size.

{\it{Peirels' substitution.}}
Magnetic field is introduced in the tight-binding model by the Peirels' substitution which introduces a phase factor for the hopping parameter:
\begin{equation}
 t_{m,n}^{m^\prime,n^\prime}\rightarrow t_{m,n}^{m^\prime,n^\prime}\mathrm{e}^{-i\frac{e}{\hbar}\int_{(m^\prime,n^\prime)}^{(m,n)}{{\bf{A}}.d{\bf{l}}}},
\end{equation}
Since the vector potential is only along the $y$-direction, the hopping parameters along the $x$-axis is unaltered (i.e. $t_x\rightarrow t_x$). The nearest neighbor hopping along y is modified as
\begin{equation}
 t_y\rightarrow t_y\mathrm{e}^{-i\frac{e}{\hbar}\int_{(m,n)}^{(m,n \pm 1)}{e{\bf{A}}.d{\bf{l}}}}=t_y\mathrm{e}^{\mp \frac{i}{\hbar}eB (ma)(b)},
\end{equation}

{\it{Hamiltonian in the magnetic unit cell.}}
We choose our magnetic unit cell such that the flux through it is in multiples of the quantum of flux.
\[
\Phi=M_x M_y B= \frac{p}{q} \Phi_0,
\]
where $M_x$ and $M_y$ are the lengths of the magnetic unit cell respectively. By convention we choose $M_y=2$ so that $M_x=q=p\frac{\Phi_0}{2B}$. So we have magnetic unit cell of length $q$. Now we proceed to write the Hamiltonian in the magnetic unit cell (which is like a unit cell with a basis of $q$ sites). The indices $r$ and $s$ denote the position of a magnetic unit cell, and the index $\nu$ denotes the site index inside a magnetic unit cell. The nearest neighbor hopping along the $x$-direction is given by the Hamiltonian,
\begin{eqnarray*}
H_x&=&-\sum_{r,s}^{}\sum_{\nu=0}^{q-2}t_{x}({c^\dagger}_{r,s,\nu}{c}_{r,s,\nu+1}+h.c) \nonumber\\
&&~~~~~~-\sum_{r,s}t_{x}({c^\dagger}_{r,s,q-1}{c}_{r+q,s,0}+h.c),
\end{eqnarray*}
and the nearest neighbour hopping along y direction by
\begin{equation}
H_y=\sum_{r,s}^{}\sum_{\nu=0}^{q-1}t_{y}({c^\dagger}_{r,s,\nu}{c}_{r,s+1,\nu}\text{e}^{il\Phi}+h.c)
\end{equation}
The total Hamiltonian is $H=H_x+H_y$.

{\it{Calculation of total energy and magnetization at zero temperature.}}
We now perform the Fourier transform defined by
\begin{equation}
c_{xy\nu}=\frac{1}{\sqrt N}\sum_{k_xk_y\in MBZ}^{}c_{k_xk_y\nu}\text{e}^{i(k_xx+k_yy)},
\end{equation}
where $MBZ$ denotes the magnetic Brillouin zone for a given magnetic field. We now define 
\begin{equation}
\psi_{\bf{k}}=\left[\begin{array}{cccc}c_{k_xk_y0}&c_{k_xk_y1}&\cdots&c_{k_xk_yq-1}\end{array}\right]
\end{equation}
Then we can write the Hamiltonian as
\begin{equation}
{H}=\sum_{(k_x,k_y)\in MBZ}^{}\psi^{\dagger}_{\bf{k}}{H(k)}\psi_{\bf{k}}
\end{equation}
where $H(k)$ is a sparse matrix with entries appropriately filled from the form of the Fourier transformed Hamiltonian.
The matrix $H(k)$ is diagonalized to obtain the band structure at a given k. The total energy at zero temperature is given by
\begin{equation}
 \epsilon_{\rm total}=\sum_{\nu,\xi^{\nu}_{\bf k}\le\mu}\xi_{\bf k}^{\nu},
\end{equation}
where $\xi_{\bf k}^{\nu}$ is the $\nu^{\rm th}$-eigenvalue of the above Hamiltonian. The magnetization at zero temperature is calculated by
\begin{equation}
M=-\frac{\partial {\epsilon_{\rm total}}}{\partial {B}}.
\end{equation}

\section{Approximate Continuum Model}
The Hamiltonian (in the absence of magnetic field) for an electron in a periodic potential can be approximated by introducing askew masses along the $x$- and $y$- directions: 
\begin{equation}
{H}={\frac{{p}_x^2}{2{m}_x}}+{\frac{{p}_y^2}{2{m}_y}}.
\end{equation}
The magnetic field applied along the $z$-direction is introduced into the Hamiltonian by the canonical replacement of the momentum operator. As in the lattice model we choose the Landau gauge where the vector potential is only along the $y$-direction given by ${\bf{A}}=Bx\hat{y}$. Vector potentials which differ by a gauge only cause a phase shift in the eigenstates.
In this choice of vector potential, the Hamiltonian becomes:
\begin{equation}
{H}={\frac{{p}_x^2}{2{m}_x}}+{\frac{({p}_y-ex{B})^2}{2{m}_y}}.
\end{equation}
This Hamiltonian commutes with ${p}_y$. Therefore we may replace ${p}_y$ with $\hbar{k}_y$, with $k_y$ taking discrete values depending on the choice of the sample size. After this substitution, the original Hamiltonian is reduced to that of a simple harmonic oscillator as
\begin{equation}
{H}={\frac{{p}_x^2}{2{m}_x}}+{\frac{(\hbar{k}_y-ex{B})^2}{2{m}_y}}.
\end{equation}
This can be written in a more standard form as:
\begin{equation}
{H}={\frac{{p}_x^2}{2{m}_x}}+{\frac{1}{2}{m}_x\left(\frac{e{B}}{\sqrt{{m}_x{m}_y}}\right)^2\left(\frac{\hbar{k}_y}{e{B}}-x_0\right)^2}.
\end{equation}
This harmonic oscillator has its center of the potential at
\begin{equation}
{{x}_0}={\frac{\hbar{k}_y}{e{B}}},
\end{equation}
and the natural frequency is
\begin{equation}
{{\omega}_0}={\frac{e{B}}{\sqrt{{m}_x{m}_y}}}.
\end{equation}
The eigenvalues of this Hamiltonian are:
\begin{equation}
{{E}_n}={\left(n+\frac{1}{2} \right)\hbar{\omega}_0}.
\end{equation}
But $k_y$ can take several values within the first Brilliuon zone, leading to degeneracy for each $n$ (called the Landau level with index $n$). We can estimate the degeneracy by the argument that the center of the potential must lie within the lattice under consideration. We consider a lattice of length $L_x$ along the $x$-direction and $L_y$ along the $y$-direction. So we obtain the condition,
\begin{equation}
0\leq{{x}_0}={\frac{\hbar{k}_y}{e{B}}}\leq{L}_x,\nonumber\\
\end{equation}
or
\begin{equation}
0\leq{{k}_y}\leq\frac{e{B}{L}_x}{\hbar}.
\end{equation}
Since in the $k$-space the number of allowed values per unit length along the $y$-direction is $\frac{2\pi}{{L}_y}$ we have that the number of allowed values $D$ is:
\begin{equation}
{D}=\frac{2e{B}{L}_x{L}_y}{h}=\frac{2e{B}S}{h},
\end{equation}
where $S$ is the area of the two dimensional lattice under consideration and the factor of $2$ accounts for the spin degeneracy. Let us assume that all the levels up to $\eta$-1 Landau levels are completely filled and the ${\eta}^{th}$ Landau level is partially filled with an occupancy of $\nu$. Thus the total number of electrons is given by:
\begin{equation}
{{N}_e}=\rho S=(\eta+ \nu)D=(\eta+\nu)\frac{2e{B}S}{h},
\end{equation}
where $\rho$ is the number of electrons per unit area. Thus,
\begin{equation}
\eta=\left[\frac{\rho h}{2q{B}}\right], \mathrm{and}\  \nu=\left\{ \frac{\rho 
h}{2q{B}_0}\right\}, 
\end{equation}
where $[ \ ]$ and $\{ \ \}$ denote the greatest integer and fractional part of the function,  respectively. Now in this configuration the total energy (the sum of energies of the Landau levels with each level having a degeneracy of $D$) is :
\begin{eqnarray}
{\epsilon_{\rm total}}&=&\left[\sum_{k=0}^{\eta-1}\left(k+\frac{1}{2} \right) + \nu \left(\eta+\frac{1}{2} \right)\right]D\hbar{\omega}_0,\nonumber\\
&=&\left(\frac{{\eta}^2}{2}+\nu \eta + \frac{\nu}{2}\right)D\hbar{\omega}_0,
\end{eqnarray}
Substituting the values for $D$ and ${\omega}_0$  we get,
\begin{equation}
{\epsilon_{\rm total}}=\frac{Sh^2\rho ^2}{2\pi \sqrt{{m}_x{m}_y}}\left(1- \frac{{\nu}(\nu-1)}{(\eta+\nu)^2} \right).
\end{equation}
{\bf {De Haas Van Alphen Oscillations.}}
The de Haas van Alphen oscillation is the oscillation of the magnetization with inverse magnetic field. The magnetization is  given by
\begin{equation}
M=-\frac{\partial {\epsilon_{total}}}{\partial {B}}=-\frac{\partial {\epsilon_{\rm total}}}{\partial (\eta+\nu)}\frac{d{(\eta+\nu)}}{d {B}},
\end{equation}
%
 \begin{equation}
{\rm where}~~~\frac{d{}}{d {B}}(\eta+\nu)=-\frac{\rho h}{2e{{B}}^2}=-\frac{2e{(\eta+\nu)}^2}{\rho h}.
\end{equation}
%
The derivative of fractional part of $\epsilon_{\rm total}$ is the same as the derivative of the original quantity unless that quantity is an integer in which case the derivative is not defined. Thus, we have
\begin{eqnarray}
\frac{\partial {\epsilon_{total}}}{\partial (\eta+\nu)}&=&\frac{Sh^2\rho ^2}{2\pi \sqrt{{m}_x{m}_y}}\nonumber\\
&&\times\left(\frac{2(\eta+\nu)\nu(\nu -1)-(r\eta+\nu)^2(2\nu -1)}{(\eta+\nu)^4} \right),\nonumber\\
\end{eqnarray}
which gives
\begin{equation}
M=\frac{2S\hbar \rho}{ \sqrt{{m}_x{m}_y}}\left(1-2\nu + \frac{2\nu(\nu -1)}{(\eta+\nu)^2} \right)
\end{equation}
%
$\nu$ always lie between 0 and 1 whereas $\eta$ is of the order of $10^8$. Therefore, the last term is extremely small compared to the other terms, and we can safely neglect it. So we have the expression for the oscillation of magnetization as
\begin{equation}
M=\frac{2S\hbar \rho}{ \sqrt{{m}_x{m}_y}}\left(1-2\nu \right)
\end{equation}
Now the fractional part of a quantity is a periodic function with period 1. Therefore the magnetization is also periodic. $\nu$ is given by $\left\{ \frac{\rho h}{2e{B}}\right\}$. This is periodic in $\frac{1}{{B}}$ with period \begin{math}\frac{2e}{\rho h}\end{math}. Now in a two dimensional lattice $\rho$ is related to the area enclosed by Fermi contour by $\rho=\frac{{S}_f}{{2\pi}^2}$, where $S_k$ is the FS area. Substituting this into the above formula for the period we obtain the familiar result:
\begin{equation}
\mathrm{\Delta\left(\frac{1}{{B}}\right)}=\frac{2\pi e}{\hbar {S}_f}.
\end{equation}
The above equation is essentially what was obtained in the Onsagar-Lifshitz calculations for closed orbit FS.

\end{document}